\documentclass{aa}
\usepackage{graphics}
\usepackage{natbib}
\bibpunct{}{}{;}{}{}{,} % to follow A&A style

\newcommand{\xmm}{{\it XMM-Newton}}

\newcommand{\rgs}{RGS}

\newcommand{\mos}{EPIC-MOS}
\newcommand{\pn}{EPIC-PN}
\newcommand{\rosat}{{\it ROSAT}}
\newcommand{\beppo}{{\it BeppoSAX}}
\newcommand{\asca}{{\it ASCA}}
\newcommand{\einstein}{{\it Einstein}}

\newcommand\lax{\>\vcenter{\hbox{$<$\hskip-.75em\lower1.0ex\hbox{$\sim$}}}\>}
\newcommand\gax{\>\vcenter{\hbox{$>$\hskip-.75em\lower1.0ex\hbox{$\sim$}}}\>}

\begin{document}

%\thesaurus{01(02.01.4;                 % atomic processes
%              02.12.1;                 % line: formation
%              02.12.2;                 % line: identification
%              03.20.8;                 % techiniques: spectroscopic
%              08.19.5 N132D;           % supernovae: individual: N132D
%              09.19.2;                 % ISM: supernova remnants
%              13.25.3)                 % X-rays: general
%}

\title{High-resolution X-ray spectroscopy and imaging of supernova remnant N132D\thanks{Based on observations obtained with \xmm, an ESA science mission with instruments and contributions directly funded by ESA Member States and USA (NASA)}}

\author{Ehud Behar\inst{1}
        \and Andrew P. Rasmussen\inst{1}
        \and R. Gareth Griffiths\inst{2}
        \and Konrad Dennerl\inst{3}
        \and Marc Audard\inst{4}
        \and Bernd Aschenbach\inst{3}
        \and Albert C. Brinkman\inst{5}         }
\titlerunning{SNR N132D observed with \xmm\ }
\authorrunning{E.Behar et al.}
\mail{behar@astro.columbia.edu}

\institute{Columbia Astrophysics Laboratory,
           550 West 120th Street, New York, NY 10027, USA
           \and X-ray Astronomy Group, Physics and Astronomy, University of 
            Leicester, Leicester LE1 7RH, U.K.
           \and Max-Planck-Institut f\"ur Extraterrestrische Physik, D-85740 Garching            , Germany
           \and Laboratory for Astrophysics, Paul Scherrer Institute, 
            W\"urenlingen and Villigen, 5232 Villigen PSI, Switzerland
           \and Space Research Organization of the Netherlands,
           Sorbonnelaan 2, 3548 CA, Utrecht, The Netherlands
            }           

\date{Received 29 September 2000 / Accepted }

\abstract{
The observation of the supernova remnant N132D by the scientific
instruments on board the \xmm\ satellite is presented. The X-rays from
N132D are dispersed into a detailed line-rich spectrum using the
Reflection Grating Spectrometers. Spectral lines of C, N, O, Ne, Mg,
Si, S, and Fe are identified. Images of the remnant, in narrow
wavelength bands, produced by the European Photon Imaging Cameras
reveal a complex spatial structure of the ionic distribution. While
K-shell Fe emission seems to originate near the centre, all of the
other ions are observed along the shell. A high O$^{6+}$ / O$^{7+}$
emission ratio is detected on the northeastern edge of the
remnant. This can be a sign of hot ionising conditions, or it can
reflect relatively cool gas. Spectral fitting of the CCD spectrum
suggests high temperatures in this region, but a detailed analysis of
the atomic processes involved in producing the O$^{6+}$ spectral lines
leads to the conclusion that the intensities of these lines alone
cannot provide a conclusive distinction between the two scenarios.   
  }

\maketitle

\section{Introduction}

N132D, one of the brightest soft X-ray sources in the Large Magellanic Cloud (LMC), was identified as a Type II supernova remnant (SNR) by Westerlund \& Mathewson\ (\cite{westerlund66}). The high velocities of the oxygen-rich (O-rich) ejecta near the centre of the remnant were first detected by Danziger \& Dennefeld\ (\cite{danziger76}) and later measured precisely by Morse et al.\ (\cite{morse95}). The fast moving ejecta is believed to originate from a core collapse supernova (SN) of a massive star. Blair et al.\ (\cite{blair00}) have recently performed a thorough optical and UV study of N132D using the Hubble Space Telescope (HST). With the exceptionally high spatial resolution of the HST, they were able to distinguish between SN ejecta and swept-up interstellar matter (ISM). Observing a knot on the northwestern rim of the remnant, they found elemental abundances that agree fairly well with those previously obtained from X-ray observations of the entire remnant (Hughes et al.\ \cite{hughes98}; Favata et al.\ \cite{favata98}), and which are also consistent with the mean LMC abundances. In the ejecta, on the other hand, only C, O, Ne, and Mg could be detected in the optical/UV spectra. The absence of O-burning elements, such as Si, S, and Fe, in the ejecta has led Blair et al.\ (\cite{blair00}) to suggest that N132D may be a product of a Type Ib SN explosion.

N132D was first observed in the X-rays with the \einstein\ Observatory (Mathewson et al.\ \cite{mathewson83}). Using the spectrometers on board \einstein, Hwang et al.\ (\cite{hwang93}) detected individual emission lines of O$^{7+}$, O$^{6+}$, Ne$^{9+}$, Ne$^{8+}$, and Fe$^{16+}$. Comparing with the \rosat\ images, Morse et al.\ (\cite{morse95}) found little overlap between the optical and soft X-ray emitting regions. The X-ray emission seemed to come from a large shell of shocked ISM gas (see also Williams et al.\ \cite{williams99}), which is also the source of strong synchrotron radio emission (Dickel \& Milne\ \cite{dickel95}). Hughes et al.\ (\cite{hughes98}) fitted the \asca\ CCD spectrum of N132D, obtaining mean LMC abundances, which further suggested that the X-rays in the remnant are predominantly emitted by shocked ISM.

In this work, we present the X-ray spectrum of N132D measured with the Reflection Grating Spectrometers (RGS), complemented by the images taken with the European Photon Imaging Cameras (EPIC), both on board \xmm. Owing to the uniquely high dispersion of the RGS, almost all of the spectral features emitted by this extended ($\sim$120\arcsec) source are resolved. Along with the RGS observation of 1E0102.2$-$7219 in the Small Magellanic Cloud (Rasmussen et al.\ \cite{andy01}), these are the first highly resolved X-ray spectra of extended SNRs.

\section{Observation and Data Reduction}

The \xmm\ X-ray observatory (Jansen et al.\ \cite{jansen01}) incorporates a payload with two identical high-resolution RGS spectrometers (den Herder et al.\ \cite{jwdh01}) and three EPIC cameras (Turner et al.\ \cite{turner01}; Str\"uder et al.\ \cite{struder01}). The spectroscopic foci are separate from the telescope foci, where the cameras are located, allowing the cameras and the spectrometers to operate simultaneously. N132D was observed by \xmm\ as part of the Performance Verification program during May, 2000. The RGS data were acquired with an effective exposure time of 53 ksec for each spectrometer. The \mos\ cameras were operated with the medium filter in place, giving effective exposure times of 18 and 21 ksec. The \pn\ camera was operated in full-frame mode using the medium filter for an effective exposure time of 23.2 ksec. 

The RGS data were processed using custom software developed at Columbia University. The background subtracted CCD events were plotted in dispersion versus pulse-height space to separate the spectral orders. An especially large (cross--dispersion) spatial extraction of 135\arcsec\ was required in order to cover the extended remnant. In order to determine the flux and provide the channel-to-wavelength transformation, the count-rate spectra were folded through the response matrices specifically generated for this observation, utilizing the ground calibration (Rasmussen et al.\ \cite{andy98}). This entire procedure was performed separately for each spectrometer and for first and second orders. Note that N132D's large angular size and velocity structure tend to broaden the spectral features irregularly, thus, preventing accurate wavelength measurements of individual lines. Fortunately, this does not impede the line identifications.

The reduction of raw \mos\ event lists, which was performed using the standard Science Analysis Software, involves the subtraction of hot, dead, or flickering pixels, as well as removal of events due to electronic noise. \mos\ images of the whole remnant were extracted using events with pattern 0 - 12 only. \pn\ data was reduced according to standard procedures. N132D covered the CCD chips 4 and 7, which required dedicated individual corrections. For the \pn\ images, areas of 44$\times$52 pixels were extracted. The extraction area was adaptively increased around each pixel to contain at least 200 single-pixel events in the energy range of 4.4 - 10 keV. In order to get reliable statistics for the spectral fits, the data were rebinned to 0.16 keV bins.

\section{Results}

\subsection{\rgs\ spectrum}

The total RGS spectrum of N132D is plotted in Fig. 1. The data from the first and second orders in both RGS have been combined in order to improve the statistical reliability. The line identifications are presented in Table 1 for the strongest lines in accordance with the peak labels in Fig. 1. For blends, only the dominant contributing lines are given in the table. Lines of C, N, O, Ne, Mg, Si, S, and Fe are unambiguously detected in the spectrum; the strongest lines originating from K-shell O and Ne, and L-shell Fe. The ratio of the strong Fe$^{16+}$ lines at $\sim$ 15 and 17 \AA\ seem to indicate that the iron has equilibrated. At least six consecutive charge states of iron, namely Fe$^{16+}$ - Fe$^{21+}$, are observed. Fe$^{22+}$ might also be present in the 11.75 \AA\ blend (label 8). This wide range of Fe charge states indicates an electron temperature ($kT_\mathrm{e}$) range of 0.2 - 1 keV in collisional equilibrium (Mazzotta et al.\ \cite{mazzotta98}). The absence of Fe$^{23+}$, which would have produced strong lines at $\sim$ 10.6 and 11.2 \AA, provides the upper limit for the temperature, whereas the lower limit is not well constrained as charge states lower than Fe$^{16+}$ do not regularly emit X-rays. However, line emission from Fe$^{24+}$ is detected with \pn\ as discussed below.

\begin{figure}
  \resizebox{\hsize}{!}{\rotatebox{0}{\includegraphics{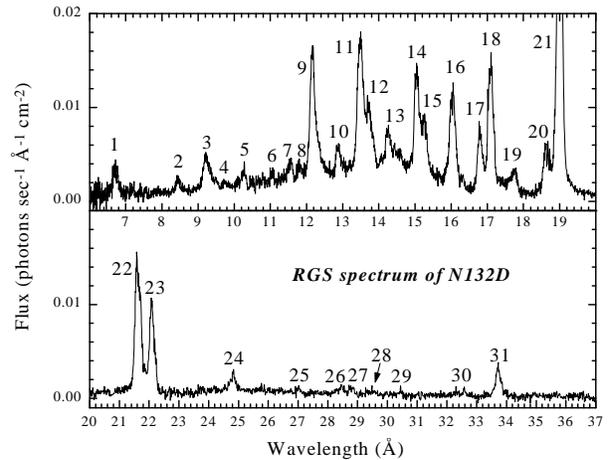}}}
  \caption{The RGS fluxed spectrum of N132D.  See Table 1 for line identifications.}
  \label{fig:f1}
\end{figure}

\begin{table}[htbp]
  \begin{center}
    \caption[Linelist]{\label{tbl:t1} Spectral lines in the RGS spectrum of N132D (Fig. 1). The references for the expected wavelengths ($\lambda_\mathrm{expected}$) are (1) Johnson \& Soff\ (\cite{johnson85}), (2) Drake \ (\cite{drake88}), (3) Brown et al.\ (\cite{brown98}) (4) Brown et al.\ (\cite{brown00}) (5) Present HULLAC (Bar-Shalom et al.\ \cite{hullac98}) calculations.}
    \begin{tabular}{rccl}
      \\ \hline \hline
      Label & $\lambda_\mathrm{expected}$ (\AA) & Reference & Ion \\ \hline \hline
      1    &  6.648 & 2 & Si$^{12+}$  \\
           &  6.685 & 2 & Si$^{12+}$  \\
           &  6.739 & 2 & Si$^{12+}$  \\
      2    &  8.421 & 1 & Mg$^{11+}$  \\
      3    &  9.169 & 2 & Mg$^{10+}$  \\
           &  9.228 & 2 & Mg$^{10+}$  \\
           &  9.314 & 2 & Mg$^{10+}$  \\
      4    &  9.708 & 5 & Ne$^{9+}$  \\
      5    & 10.238 & 5 & Ne$^{9+}$  \\
      6    & 11.009 & 5 & Ne$^{8+}$  \\
      7    & 11.554 & 5 & Ne$^{8+}$  \\
      8    & 11.742 & 4 & Fe$^{22+}$  \\
           & 11.770 & 4 & Fe$^{21+}$  \\
      9    & 12.135 & 1 & Ne$^{9+}$  \\
     10    & 12.846 & 4 & Fe$^{19+}$  \\
           & 12.864 & 4 & Fe$^{19+}$  \\
     11    & 13.447 & 2 & Ne$^{8+}$  \\
           & 13.518 & 4 & Fe$^{18+}$  \\
           & 13.550 & 2 & Ne$^{8+}$  \\
     12    & 13.698 & 2 & Ne$^{8+}$  \\
     13    & 14.208 & 4 & Fe$^{17+}$  \\
     14    & 15.015 & 3 & Fe$^{16+}$  \\
     15    & 15.261 & 3 & Fe$^{16+}$  \\
     16    & 16.005 & 1 & O$^{7+}$  \\
           & 16.004 & 4 & Fe$^{17+}$  \\
     17    & 16.779 & 3 & Fe$^{16+}$  \\
     18    & 17.051 & 3 & Fe$^{16+}$  \\
           & 17.096 & 3 & Fe$^{16+}$  \\
     19    & 17.623 & 4 & Fe$^{17+}$  \\
     20    & 18.627 & 5 & O$^{6+}$  \\
     21    & 18.967 & 1 & O$^{7+}$  \\
     22    & 21.602 & 2 & O$^{6+}$  \\
           & 21.804 & 2 & O$^{6+}$  \\
     23    & 22.097 & 2 & O$^{6+}$  \\
     24    & 24.779 & 1 & N$^{6+}$  \\
     25    & 26.988 & 5 & C$^{5+}$  \\
     26    & 28.464 & 5 & C$^{5+}$  \\
     27    & 28.780 & 2 & N$^{5+}$  \\
           & 29.082 & 2 & N$^{5+}$  \\
     28    & 29.535 & 2 & N$^{5+}$  \\
     29    & 30.436 & 5 & S$^{13+}$  \\
     30    & 32.436 & 5 & S$^{13+}$  \\
           & 32.579 & 5 & S$^{13+}$  \\
     31    & 33.734 & 1 & C$^{5+}$   \\
  \\ \hline \hline
    \end{tabular}\\[1.0ex]
  \end{center}
\end{table}

Modeling the profiles and fluxes of the individual spectral lines in the RGS spectrum is a rather difficult task. The line profiles are correlated with the shape of the remnant, the velocity broadening, and the point spread function of the instrument (telescope + spectrometer). The fact that the former two effects may strongly depend on the ion species makes this problem especially challenging. Therefore, with the sole exception of the O$^{6+}$ lines discussed below, a detailed line-flux analysis is deferred to a later study.

\subsection{\mos\ soft X-ray images}

Fig. 2 shows the \mos\ images of N132D in several narrow wavelength bands that overlap with those of the RGS. The data (3\arcsec$\times$3\arcsec\ pixels) in each band are combined from the two cameras, smoothed using a Gaussian function (sigma=1.5\arcsec), and then normalised to the brightest pixel in each image. Each narrow-band image is 2\arcmin$\times$2.5\arcmin\ wide. All of the images feature strong X-rays from the shell, in particular from the southeastern, and northwestern edges. Differences between more and less ionised regions can be noticed. The emission by the most highly ionised species Ne$^{9+}$, Mg$^{10+}$, Si$^{12+}$, and Fe$^{19+}$ is brightest on the southeastern tip, while O$^{6+}$, O$^{7+}$, Ne$^{8+}$, Fe$^{16+}$, and Fe$^{17+}$ exhibit strong emission on the northwestern rim as well. This is the ISM shocked region observed by Blair et al.\ (\cite{blair00}). O$^{7+}$ emission is also intense at the centre of the remnant, indicating that part of the X-ray flux may be associated with the SN ejecta. The O$^{6+}$ image is unique by virtue of the bright knot on the northeastern edge, just between the two broad inlets that extend into the ambient ISM. In fact, no other element seems to be emitting X-rays in that region. The \mos\ ratio image of O$^{6+}$ / O$^{7+}$ shown in Fig. 3 illustrates the excess of O$^{6+}$ emission on the northeastern rim, where the ratio attains a value of 0.79$\pm0.01$ in contrast with 0.60$\pm0.01$ in the centre. The RGS dispersed images for these lines (not presented) are consistent with this picture. The O$^{6+}$ brightness might simply be due to relatively cold gas. However, the special morphology in that region may, alternatively, suggest that the shock wave is interacting with relatively dense and cold O-rich material, possibly from pre-SN stellar winds. In that case, this newly shocked material can be very hot and ionising. It should be noted that Banas et al.\ (\cite{banas97}) have observed a giant molecular cloud near the {\it southern} edge of N132D, whereas no clouds were detected to the north of the remnant.

\begin{figure}
  \resizebox{\hsize}{!}{\rotatebox{0}{\includegraphics{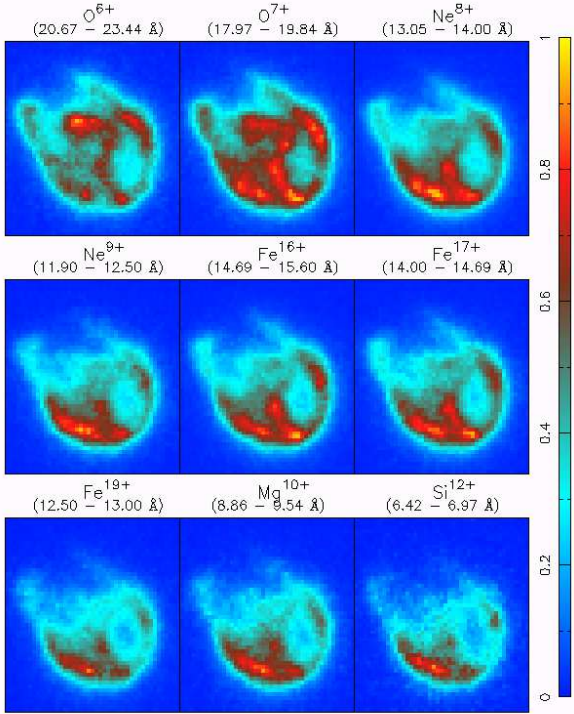}}}
  \caption{\mos\ images of N132D in the narrow wavelength bands indicated. Each image is labeled with the principal line-emitting ion in its particular band.}
  \label{fig:f2}
\end{figure}

\begin{figure}
  \resizebox{\hsize}{!}{\rotatebox{0}{\includegraphics{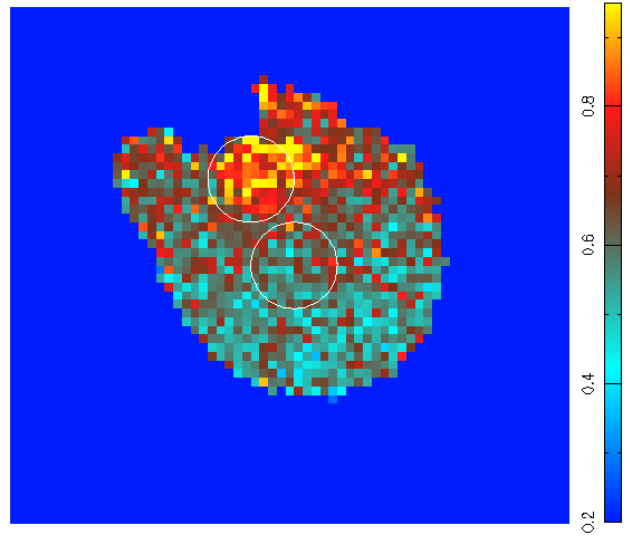}}}
  \caption{\mos\ O$^{6+}$ / O$^{7+}$ ratio image of N132D. The average ratios in the circular regions (15\arcsec radius) are 0.79$\pm0.01$ near the northeastern rim and 0.60$\pm0.01$ in the centre.}
  \label{fig:f3}
\end{figure}

\subsection{\pn\ hard X-ray images}
The \pn\ camera is more sensitive to hard X-rays than the RGS and \mos. The \pn\ CCD spectrum of the entire N132D remnant (not presented) shows K-shell features of Ar and Ca, as well as a prominent Fe$^{24+}$ feature at 1.86 \AA\ (6.67$\pm0.01$ keV). This Fe-K feature was also observed with \beppo\ (Favata et al.\ \cite{favata98}). In Fig. 4(a), flux images of N132D in this line (blend) are presented. In order to separate continuum from line flux, each spectrum was fitted with a Gaussian line superimposed on a bremsstrahlung continuum. 

The N132D structure as seen in the \pn\ images (Fig. 4) is completely different from the structure in the \mos\ images (Fig. 2). Unlike the softer emitting material that exhibits a shell with some filaments, the harder Fe-K line originates in a well-defined region slightly offset from the centre of the remnant to the southeast, where there is no sign for Fe-L emission. This is consistent with an out moving blast wave; the material closer to the centre that was shocked earlier is more highly ionised. If this material is associated with the SN ejecta, it might suggest that the high-Z elements are not absent from the ejecta, but rather that they could not be detected at shorter wavelengths because they are too ionised. The northwest and northeast regions show almost no Fe$^{24+}$ emission. The continuum flux is highest near the centre (Fig. 4c), which makes N132D look like a centre-filled remnant, rather than a shell-type remnant, as suggested by the soft X-ray images. The highest equivalent width (EW) values of more than a few keV are found in a confined northern region, where the Fe abundance is probably high and the overall density low. The hottest regions in N132D are found close to the northern inlets and are devoid of any Fe-K line emission, but do exhibit some continuum flux. This can be either due to depletion of iron in those regions, but can also be explained by the temperature exceeding 10 keV or by a non-thermal continuum. Note that the bright O$^{6+}$ spot (Fig. 2) seems to lie close to these hot regions, in which case O$^{6+}$ would be definitely ionising.

\begin{figure}
  \resizebox{\hsize}{!}{\rotatebox{0}{\includegraphics{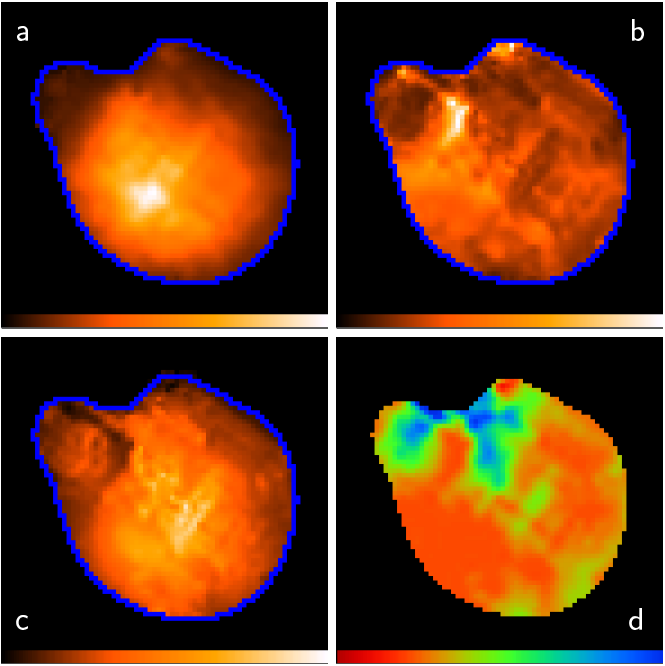}}}
  \caption{N132D 150\arcsec$\times$150\arcsec\ maps of (a) the Fe-K line flux at 6.67 keV, (range 0 - 1.66 10$\sp{-5}$ photons cm$\sp{-2}$ s$\sp{-1}$ arcmin$\sp{-2}$), (b) the EW of the Fe-K line, (0 - 4.7 keV), (c) the differential continuum flux at 6.67 keV derived from the 4.4 - 10 keV spectrum, (0 - 1.15 10$\sp{-5}$ photons cm$\sp{-2}$ s$\sp{-1}$ keV$\sp{-1}$ arcmin$\sp{-2}$), (d) the {\it apparent} temperature (0.4 - 3.1 keV) derived from the continuum only. All colour bars are linear.} 
  \label{fig:f4}
\end{figure}

In order to understand the absence of Fe$^{23+}$ in the RGS spectrum and the weakness of Fe$^{22+}$ and Fe$^{21+}$, while emission from both lower and higher charge states is intense, the entire \pn\ spectrum above 2.5 keV has been fitted with a two-temperature model. An acceptable fit is obtained for $kT_e$= 0.89 and 6.2 keV. These results can explain the absence of Fe$^{23+}$, which attains its maximal fractional abundance at the intermediate temperature of  $\sim$1.5 keV (Mazzotta et al.\ \cite{mazzotta98}). The absence of these intermediate temperatures most likely means that the hot and cold regions are spatially distinct. 

\section{Spectroscopic analysis of the O$^{6+}$ lines}

Seeking further evidence that the O-rich gas in the northeastern part of N132D is in the process of being ionised, we examine the RGS spectrum in the vicinity of the three strong emission lines of O$^{6+}$ (He-like triplet). These lines are at 21.602, 21.804, and 22.097 \AA\ and are referred to, respectively, as the resonant (r), intercombination (i), and forbidden (f) lines. The i line is very weak in the N132D spectrum, whereas the r and f lines are strong, as can be seen in Fig. 5. A hint for similar line structure is found for the N$^{5+}$ triplet as well (Fig. 5). Fitting the O$^{6+}$ spectrum with the most appropriate line-profile model available yields line intensity ratios of f : i : r = 0.62 : 0.12 : 1.0. However, since the model can not very well reproduce the observed line shapes, uncertainties of about $\pm$20\% should be associated with these values. The quality of the data for N$^{5+}$ is insufficient to allow for a similar measurement with the N$^{5+}$ lines.

Decaux et al. (1997) pointed out that for He-like Fe, intense r and f lines together with a very weak i line is an indication for hot ionising conditions, because: First, at high temperature the r line is enhanced compared to the i and f lines. Second, inner-shell ionisation processes from the ground state 1s$^2$2s of the Li-like ion to the 1s2s (J=1) level enhance the f line. In the present case of O and N, inner-shell ionisation of Li-like ions also produces He-like excited ions in the 1s2s configuration. This configuration has two levels; the J=1 level is the upper level of the f line, whereas the J=0 level cannot decay radiatively to the ground level and in the case of Fe cannot decay radiatively at all. In contrast with the Fe case, for O and N, the 1s2p$_{1/2}$ (J=1) level, which is the upper level of the i line, lies {\it below} the 1s2s (J=0) level. This allows for radiative decay that enhances the i line in parallel to the f-line enhancement.

\begin{figure}
  \resizebox{\hsize}{!}{\rotatebox{0}{\includegraphics{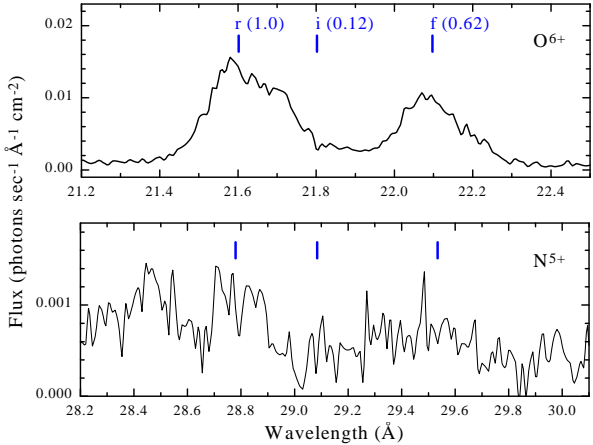}}}
  \caption{The RGS spectrum of N132D in the limited wavelength range corresponding to the He-like triplets of O$^{6+}$ and N$^{5+}$ ($\lambda_{expected}$ indicated by horizontal bars). Relative line intensities are given for O$^{6+}$. }
  \label{fig:f5}
\end{figure}

We use atomic-state models employing data from the HULLAC code (Bar-Shalom et al.\ \cite{hullac98}) to calculate the spectrum of O$^{6+}$ in a wide range of temperatures, both in ionising plasma and in equilibrium. Indeed, it is found that both situations can produce the observed f : i : r ratios. Both hot (e.g., 0.6 keV) ionising gas with a rather low O$^{6+}$ / O$^{5+}$ ratio (e.g., 2), as well as a cooler (0.15 keV) gas in equilibrium can emit these lines with the observed ratios. In other words, the intensities of the O$^{6+}$ lines alone are insufficient for distinguishing between the equilibrium and ionising scenarios. For completeness, it should be noted that the current spectrum at least rules out a cooling plasma for N132D, because in that case, the upper levels of the i and f lines are populated by recombination resulting in much higher f/r and i/r ratios than are observed here.

\section{Conclusions}

The RGS spectrum of N132D features lines of C, N, O, Ne, Mg, Si, S, and Fe. K-shell Ar, Ca, and Fe are detected in the \pn\ spectrum. The narrow-band EPIC images of N132D give a coarse mapping of the elemental and temperature structure in the remnant. With the exception of O, the dominant part of the soft X-ray emission originates from shocked ISM along the southeastern, and northwestern edges of the expanding shell. In contrast, strong Fe-K emission is detected near the centre of the remnant, perhaps indicating that high-Z ejecta is too highly ionised to be observed in shorter wavelengths. O$^{7+}$ is present throughout the remnant, while the brightest spot of O$^{6+}$ emission is near the northeastern rim. The O$^{6+}$ emission can be attributed, either to low temperatures, or to hot, recently shocked material that is in the process of being ionised. Fitting of the \pn\ spectrum indicates high temperatures in that region. Unfortunately, the emission lines of O$^{6+}$ alone, although clearly seen in the RGS spectrum, cannot provide a conclusive distinction between these two scenarios, due to considerations associated with the atomic processes involved in producing these lines. A more detailed analysis of the RGS spectrum, and especially of the many strong Fe lines, will potentially provide a better insight into the temperature and non-equilibrium structure of the hot gas in N132D.


\begin{thebibliography}{}

  \bibitem[1997]{banas97} Banas K. R., Hughes J. P., Bronfman L., 
    Nyman L.-\AA\, 1995 ApJ, 480, 607
  \bibitem[1998]{hullac98} Bar-Shalom A., Klapisch M., Goldstein W. H., Oreg J., 
    1998 {\it The HULLAC code for atomic physics}, (unpublished).
  \bibitem[2000]{blair00} Blair W. P., Morse J. A., Raymond J. C., et al.,
%    Kirshner R. P., Hughes J. P., Dopita M. A., Sutherland R. S., Long K. S., 
%    Winkler P. F., 
    2000 ApJ, 537, 667
  \bibitem[1998]{brown98} Brown G. V., Beiersdorfer P., Liedahl D. A., Widmann K., 
    Kahn S. M., 1998 ApJ, 502, 1015
  \bibitem[2000]{brown00} Brown G. V., Beiersdorfer P., Liedahl D. A., Widmann K., 
    Kahn S. M., 2000 LLNL preprint (UCRL-JC-136647)
  \bibitem[1976]{danziger76} Danziger I. J., Dennefeld M., 1976
    ApJ, 207, 394
  \bibitem[1997]{decaux97} Decaux V., Beiersdorfer P., Kahn S. M., 
    Jacobs V. L., 1997 ApJ, 482, 1076
  \bibitem[2001]{jwdh01} den Herder J. W., Brinkman A. C, Kahn S. M., et al., 
    2001 A\&A, 365 (this issue)
  \bibitem[1988]{drake88} Drake G. W., 1988 Can. J. Phys., 66, 586
  \bibitem[1995]{dickel95} Dickel J. R., Milne D. K., 1995 AJ, 
    109, 200
  \bibitem[1998]{favata98} Favata F., Vink J., Parmar A. N., Kaastra J., Mineo T., 
    1998 A\&A 340, 626
%  \bibitem[2000]{flanagan00} Flanagan K. A., Canizares C. R., Davis D. S., 
%    Dewey D., Houck J. C., Schattenburg M. L., 2000 BAAS 196, 3409
  \bibitem[1998]{hughes98} Hughes J. P., Hayashi I., Koyama K., 1998
    ApJ, 505, 732
  \bibitem[1993]{hwang93} Hwang U., Hughes J. P., Canizares C. R., 
    Market T. H., 1993 ApJ, 414, 219
  \bibitem[2001]{jansen01} Jansen, F., Lumb, D., Altieri, B.\ et al.\ 2001 A\&A, 365
    (this issue)
  \bibitem[1985]{johnson85} Johnson W. R., Soff G., 1985 Atom. Data Nucl. Data Tables,     33, 405
  \bibitem[1983]{mathewson83} Mathewson D. S., Ford V. L., Dopita M. A.,
    Tuohy I. R., Long K. S., Helfand D. J., 1983 ApJS, 51, 345
  \bibitem[1998]{mazzotta98} Mazzotta P., Mazzitelli G., Colafrancesco S., 
    Vittorio N., 1998 A\&AS, 133, 403
  \bibitem[1995]{morse95} Morse J. A., Winkler P., Kirshner R. P., 1995 AJ, 
    109, 2104
%  \bibitem[1996]{morse96} Morse J. A., Blair W. P., Dopita M. A., Hughes J. P.,
%    Kirshner R. P., Long K. S., Raymond J. C., Sutherland R. S., 
%    Winkler P., 1996 AJ 112, 509
  \bibitem[1998]{andy98} Rasmussen A., Cottam J., Decker T., et al., 1998, 
    SPIE 3444, 327  
%  \bibitem[1998b]{andy98b} Rasmussen A., et al., 1998, in: Dahlem M. (ed.), 
%    Science with XMM: Proceedings of the First XMM Workshop
  \bibitem[2001]{andy01} Rasmussen A., Behar E., Kahn S. M. van der Heyden K. 
    , den Herder J. W., 2001 A\&A, 365 (this issue)
%  \bibitem[1990]{russell90} Russell S. C., Dopita M. A., 1990,
%    ApJS 74, 93
  \bibitem[2001]{struder01} Str\"uder, L., Briel, U. G., Dennerl, K.\ et al.\ 2001
    A\&A, 365 (this issue)
  \bibitem[2001]{turner01} Turner, M. J. L., Abbey, A., Arnaud, M.\ et al.\ 2001
    A\&A, 365 (this issue)
%  \bibitem[1992]{vancura92} Vancura O., Blair W. P., Long K. S., 
%    Raymond J. C., 1992, ApJ 394, 158
  \bibitem[1966]{westerlund66} Westerlund B. E., Mathewson D. S., 1966,
    MNRAS 131, 371
  \bibitem[1999]{williams99} Williams R. M., Chu Y. H., Dickel J. R., Petre R.,
    Smith R. C., Tavarez M., 1999, ApJS 123, 467
\end{thebibliography}
\end{document}